\documentclass[twocolumn,useAMS,usenatbib]{mnras} \usepackage{natbib}
\usepackage{amsmath,latexsym,amssymb} \usepackage{epsfig,graphicx}
\usepackage[draft]{todonotes}

\usepackage{color,soul,array,comment}

 \def\reff@jnl#1{{\rm#1\/}}

\newcommand{\beq}{\begin{equation}} \newcommand{\eeq}{\end{equation}}
\newcommand{\beqa}{\begin{eqnarray}} \newcommand{\eeqa}{\end{eqnarray}}
\newcommand{\code}[1]{{\sc #1}}

\newcommand{\galsim}{{\sc GalSim}}
\newcommand\sersic{S\'{e}rsic}

\newcommand\noteag[1]{\todo[color=yellow, inline, size=\small]{AG: #1}}

\newcommand\refresp[1]{#1}

\newcommand\rmxb[2]{\ensuremath{y=#1 \ensuremath{r_{1/2}}' + #2}}
\newcommand\nmxb[2]{\ensuremath{y=#1 \ensuremath{n}' + #2}}
\newcommand\nuu[1]{\ensuremath{\nu_\text{#1}}}

\title[Correlated noise]{The impact of correlated noise on galaxy shape estimation for weak lensing}

\author[Gurvich \& Mandelbaum] 
{Alex Gurvich$^1$, Rachel Mandelbaum$^1$\thanks{\tt rmandelb@andrew.cmu.edu} \\
$^1$McWilliams Center for Cosmology, Department of Physics, 
Carnegie Mellon University, Pittsburgh, PA 15213, USA
}

\date{\today}

\begin{document} \maketitle

\begin{abstract} 
  The robust estimation of the tiny distortions (shears) of galaxy shapes caused by weak gravitational lensing in the presence
  of much larger shape distortions due to the point-spread function (PSF) has been widely investigated.
  One major problem is that most galaxy shape measurement methods are subject to bias due to pixel
  noise in the images (``noise bias'').  Noise bias is usually 
  characterized using uncorrelated noise fields; however, real images typically have low-level
  noise correlations 
  due to  galaxies below the detection threshold, and some types of image processing can induce
  further noise correlations. We investigate the effective
  detection significance and its impact on noise bias in the presence of correlated noise for one
  method of galaxy shape estimation. 
  For a fixed noise variance, the biases in galaxy shape estimates
  can differ substantially for uncorrelated versus correlated noise.  However, use of an
  estimate of detection significance that accounts for the noise correlations can almost 
  entirely remove these differences, leading to consistent values of noise bias as a
  function of detection significance for correlated and uncorrelated noise.  We confirm the
  robustness of this finding to properties of the galaxy, the PSF, and the noise field, and quantify the
  impact of anisotropy in the noise correlations.  Our results highlight the importance of
  understanding the pixel noise model and its impact 
  on detection significances when correcting for noise bias on weak lensing.
\end{abstract}



\section{Introduction}\label{sec:intro}

Gravitational lensing, the deflection of light by mass, has been used for many different 
applications due to its sensitivity to all gravitating mass, including dark matter.  Weak
gravitational lensing \citep[for a review,
see][]{2001PhR...340..291B,2003ARA&A..41..645R,schneider06,2008ARNPS..58...99H,2010RPPh...73h6901M,2013PhR...530...87W},
which causes tiny but coherent distortions in galaxy shapes, can reveal the dark matter halos in
which galaxies live
\citep[e.g.,][]{2012ApJ...744..159L,2013ApJ...778...93T,2014MNRAS.437.2111V,2015MNRAS.449.1352C,2015MNRAS.446.1356H,2015MNRAS.447..298H,2015MNRAS.454.1161Z},
can constrain the 
amplitude of matter fluctuations and reveal the impact of dark energy on structure growth 
\citep[e.g.,][]{2013MNRAS.432.2433H,2013ApJ...765...74J,2013MNRAS.432.1544M}, and teach us about the 
theory of gravity on cosmological scales \citep[e.g.,][]{2010Natur.464..256R,2013MNRAS.429.2249S,2015MNRAS.449.4326P}.

As lensing surveys grow and their statistical errors decrease in size, 
systematic errors become more important.  Among the main systematic errors in weak lensing measurements are
those that arise when trying to accurately infer the shears (shape distortions) based on
galaxy shape measurements 
\citep[e.g.,][]{2006MNRAS.368.1323H,2007MNRAS.376...13M,2010MNRAS.405.2044B,2012MNRAS.423.3163K,2015MNRAS.450.2963M}. 
An important component of this problem 
is the substantial bias that arises 
due to pixel noise (``noise bias'')
\citep{2004MNRAS.353..529H,2012MNRAS.427.2711K,2012MNRAS.424.2757M,2012MNRAS.425.1951R}, which is typically larger than
the statistical uncertainties expected from weak lensing datasets in the next few years.

The noise model in astronomical images is typically a combination of Poisson noise on the pixel
counts plus other sources of noise that are Gaussian (e.g., read noise).  The noise can commonly be
modeled as stationary on the scale of galaxy images, at least for the faint galaxies that dominate
weak lensing measurements.   In this regime, the variance is the same in each pixel,
which is true when the Poisson noise on the sky level dominates and when the sky level does not vary
much across each galaxy.  In an unprocessed image, the noise is largely uncorrelated between pixels, though the
light profiles of galaxies that are just below the detection threshold do induce some 
noise correlations.  Even more significant 
correlations between the noise in different pixels can be induced via processes such as correction
for charge transfer inefficiency \citep{2010MNRAS.401..371M} and image resampling 
\citep{1999PASP..111..227L,2002PASP..114..144F,2011PASP..123..497F,2011ApJ...741...46R} as part of the image
combination process, particularly if the images are resampled to smaller pixels than those on the
original detector.  In combined images from the {\em HST} (Hubble Space Telescope) used for weak lensing, this effect can
be quite significant \citep[e.g.,][]{2007ApJS..172..196K}, enough so that naive estimates of
signal-to-noise ratios are wrong by 
factors of more than two if the correlations are ignored \citep{2000AJ....120.2747C}.

In this paper, we therefore consider the issue of noise bias in the context of correlated noise.  
\refresp{There are several (essentially interchangeable) ways to think about noise bias.  In
  general, it can be thought of as a change in shape of the likelihood surface due to noise \citep{2012MNRAS.425.1951R}, such
  that a maximum 
  likelihood estimator of the per-object shape is a biased estimator.  Alternatively, for the case
  of moments-based estimators, the bias arises because PSF correction involves the division of two
  noisy quantities, which has an expectation value that is not mathematically equivalent to the
  ratio of their (noiseless) 
  expectation values. The size of the bias can be written as a Taylor expansion in the inverse detection
  significance \citep[e.g.,][]{2004MNRAS.353..529H}.} 

\refresp{In light of this background, the following questions are of 
particular interest.}  First, does noise bias have any fundamentally
different characteristics in the presence of correlated noise?  \refresp{Given the mathematical
  origin of the effect, it seems that the noise bias should be described the same way for
  uncorrelated and correlated noise, provided that the detection significance is properly quantified.
  Given this requirement, the second question is whether one can use} simulations with
uncorrelated noise to calibrate noise bias \refresp{even for data with correlated noise}
after accounting for the way correlated noise modifies the signal-to-noise ratio (SNR) of the object
detections?  Third, should such corrections use the SNR of the galaxy size, flux, or shape?  
Finally, if correlated noise has some directionality, how large is the expected bias
\citep{2012MNRAS.420.1518M} in shear estimates?    We address these questions using
\galsim\ \citep{2015A&C....10..121R}, an
open-source\footnote{\url{https://github.com/GalSim-developers/GalSim}} image simulation 
package that was designed for tests of weak lensing measurement algorithms.  It includes routines to
simulate data with  uncorrelated or correlated noise with a particular correlation function, making
it ideal for this application.  

In case the answers to these questions depend on the galaxy shape measurement 
method, we use two quite different methods.  The first is a moment-based routines
(re-Gaussianization; \citealt{2003MNRAS.343..459H}) used for the majority of this work, but a subset
of our measurements use a forward model-fitting
method, \code{im3shape} 
\citep{2013MNRAS.434.1604Z,2014ascl.soft09013Z}.  
A comparison of results with these methods may give some insight into the general
applicability of our results.

The outline of this paper is as follows.  Sec.~\ref{sec:background} defines 
the quantities discussed in the rest of the paper.  Sec.~\ref{sec:methods}
includes a description of the simulations used for the noise bias investigations.  Our results are
in Sec.~\ref{sec:results}, with a discussion of the implications for a real data analysis in Sec.~\ref{sec:conclusions}.

\section{Technical Background}\label{sec:background}

In this section, we define the quantities that are commonly measured from galaxy
images used for weak lensing, along with the uncertainty on those quantities.

\subsection{Galaxy properties}\label{subsec:properties}

\subsubsection{Galaxy shape}

The majority of the methods of estimating weak lensing shear distortions from imaging data involve a two-step
process \citep[see, e.g.,][]{2015MNRAS.450.2963M}.  The first step is to estimate a per-galaxy shape
(defined in one of several ways), and the second step is to combine all the per-galaxy shapes to
estimate the cosmological shear or its correlation function for the ensemble.  The prevalence of the
first step in methods that have been used for realistic weak lensing analysis\footnote{For
  alternatives that are under development, see, e.g., \cite{2014MNRAS.438.1880B} and \cite{2015JCAP...01..024Z}.} motivates our
focus on per-galaxy shape estimates as an intermediate step towards the end goal of inferring
lensing shear.

Weak gravitational lensing can be described as a linear
transformation between unlensed and lensed coordinates, as encoded in 
the two components of the complex-valued lensing shear
$\gamma = \gamma_1 + {\rm i}\gamma_2$ and the lensing convergence $\kappa$.  The
shear describes the \emph{stretching} of galaxy images due to
lensing, while the convergence
$\kappa$ describes a change in apparent size and flux for lensed objects at fixed surface
brightness.  
In practice, the observed quantity is the reduced shear, $g_i=\gamma_i/(1-\kappa)$.

Estimation of shear typically starts with some measure of the galaxy shape.
Despite the fact that galaxies do not in general have elliptical isophotes, galaxy light profiles
are typically modelled as having  a 
well-defined ellipticity, $\varepsilon = (\varepsilon_1, \varepsilon_2)$, with
magnitude $|\varepsilon| = \sqrt{\varepsilon_1^2 + \varepsilon_2^2} = (1 - b/a) / (1 + b/a)$, where
$b/a$ is the semi-minor to semi-major axis ratio, and orientation 
angle $\phi$ of  the major axis with respect to some fixed coordinate system.  The two shape components can be
defined as $\varepsilon_1 = |\varepsilon|\cos{(2\phi)}$ and $\varepsilon_2 =
|\varepsilon|\sin{(2\phi)}$, or combined as a complex ellipticity $\varepsilon = \varepsilon_1 +
{\rm i}\varepsilon_2$.  
 For a randomly-oriented population of source ellipticities,
the ensemble average ellipticity after lensing
is an unbiased estimate of the reduced shear: $\langle \varepsilon \rangle
\simeq g $.

Another common choice of shape parametrization is based on second moments of the
galaxy image,
\begin{equation}\label{eq:qij}
Q_{ij} = \frac{\int {\rm d}^2 x I({\bf x}) W({\bf x}) x_i x_j }
{\int {\rm d}^2 x I({\bf x}) W({\bf x}) },
\end{equation}
where the coordinates $x_1$ and $x_2$ correspond to the $x$ and $y$
directions (respectively), $I({\bf x})$ denotes the galaxy image light
profile, $W({\bf x})$ is a 
weighting function (see \citealp{schneider06}), and 
the coordinate origin ${\bf x} = 0$ is at the galaxy
image center (the centroid). A second definition
of ellipticity, sometimes referred to as the \emph{distortion}, can be written as
\begin{equation}\label{eq:ellipticity}
e = e_1 + {\rm i} e_2 = \frac{Q_{11} - Q_{22} + 2 {\rm i} Q_{12}}{Q_{11} + Q_{22}}.
\end{equation}

For many weight functions $W$,  an image with
elliptical isophotes of axis ratio $b/a$ has
\begin{equation}
|e| = \frac{1 - b^2 / a^2}{1 + b^2 / a^2}.
\end{equation}
For a randomly-oriented population of source distortions,
the ensemble average $e$ after lensing
gives an unbiased estimate of approximately twice the shear that depends on the
population root mean square (RMS) ellipticity, $\langle e \rangle
\simeq 2[1-\langle (e^{(s)})^2\rangle] g $.

Finally, in the Gaussian approximation \citep{2002AJ....123..583B,2012MNRAS.425.1951R}, the statistical uncertainty on
$e$ and $\varepsilon$ are $\sigma_e=2(1-e^2)/\nu\approx 2/\nu$ and
$\sigma_\varepsilon=(1-\varepsilon^2)/\nu\approx 1/\nu$. Here $\nu$ is a total
detection significance.  Using simulations with many noise realizations to estimate the
uncertainty on $e$ allows for the definition of a shape-based detection-significance
$\nu_\text{e}$.  In practice, we use the approximate relations above (\refresp{$\nu_e\equiv 2/\sigma_e$,} without the factor of $1-e^2$)
since per-object scatter in $e$ would result in very uncertain $\nu$ estimates; however, 
when using the true value of $e$, the conclusions of the paper are unchanged.

\subsubsection{Galaxy shape estimation methods}\label{subsec:shapemeas}

Our work relies on two methods of galaxy shape estimation, re-Gaussianization
\citep{2003MNRAS.343..459H} and \code{im3shape} 
\citep{2013MNRAS.434.1604Z,2014ascl.soft09013Z}, that use a distortion $e$ and ellipticity
$\varepsilon$ (respectively) to characterize shapes.

The re-Gaussianization method is a modified version of ones that use ``adaptive moments'' (which are
equivalent to fitting the light intensity profile to an elliptical Gaussian).  The
re-Gaussianization method involves determining shapes of the PSF-convolved galaxy image
based on adaptive moments and then correcting the resulting shapes based on adaptive moments of the
PSF.  There are also additional steps to correct for non-Gaussianity of both the PSF and the galaxy
surface brightness profiles \citep{2003MNRAS.343..459H}.

\code{im3shape} is a maximum-likelihood fitting method.  It iteratively fits the selected type of
galaxy model to the observed galaxy image, using an oversampled image of the PSF.  Our adopted
settings for all \code{im3shape} calculations in this work are given in Appendix~\ref{app:im3shape}.

\subsubsection{Galaxy size}

Many forward-modeling methods of estimating galaxy shapes also include an estimate of the intrinsic 
galaxy size \citep[before convolution with the PSF; see, e.g.,][]{2013MNRAS.434.1604Z}.
Estimates of galaxy intrinsic size are also useful for estimates of lensing magnification that
take advantage of size information \citep{2012ApJ...744L..22S,2014ApJ...780L..16H}.  
With \code{im3shape}, the natural size estimate to use is the intrinsic half-light radius, $r_{1/2}$.

While measurements of the observed (PSF-convolved) size $\sigma$ from the elliptical Gaussian-weighted
adaptive moments \citep{2003MNRAS.343..459H} do not say much about the real galaxy light profile,
they can be useful for quantifying the detection significance:
\beq
\nu_\text{size} = \frac{\sigma}{\text{Uncertainty on }\sigma}.
\eeq

\subsubsection{Flux} 

Another type of detection significance is derived from the flux from the adaptive moments, and its
uncertainty.  Note that the adaptive moments 
flux may differ from the total flux due to PSF-convolved galaxies not
having Gaussian profiles.  Nonetheless, it is a fast and efficient way to estimate another form of
detection significance:
\beq
\nu_\text{flux} = \frac{\text{flux}}{\text{Uncertainty on flux}}.
\eeq

\subsection{SNR estimators}

Estimates of noise bias using simulations are typically done as a function of the SNR of the
object detection.  Several weak lensing community challenges
\citep{2009AnApS...3....6B,2014ApJS..212....5M} have used an optimal SNR estimator that 
uses the object light profile itself as a weight function, which is only possible if
the true noise-free light profile is known.  This optimal estimator, $\nu_\text{ideal}$, is defined as
follows.  The signal $S$ can be defined using a weighted sum over the pixels in the image,
\begin{equation}\label{eq:signal}
S = \frac{\sum W({\bf x}) I({\bf x})}{\sum W({\bf x})},
\end{equation}
and its variance is
\begin{equation}\label{eq:signalvar}
\text{Var}(S) = \frac{\sum W^2({\bf x}) \text{Var}(I({\bf x}))}{(\sum W({\bf x}))^2}.
\end{equation}
In the limit that the sky background dominates, $\text{Var}(I({\bf x}))$
is a constant, denoted $\text{Var}(I({\bf x}))=\sigma_\text{p}^2$ 
(the pixel noise variance).  Adopting a matched filter
for $W$, i.e., $W({\bf x})=I({\bf x})$ and putting our assumptions into
Eqs.~\eqref{eq:signal} and~\eqref{eq:signalvar} gives
\begin{equation}\label{eq:sndef}
\nu_\text{ideal} = \frac{\sqrt{\sum I^2({\bf x})}}{\sigma_\text{p}}.
\end{equation}

However, with noisy images, $\nu_\text{ideal}$ is not a measurable quantity, forcing us to use 
realistic estimators of SNR that can differ from $\nu_\text{ideal}$ by as much as a factor
of two \citep[see, e.g.,][]{2015MNRAS.450.2963M}.  As noted in Sec.~\ref{subsec:properties}, there are several possible
empirical estimates of 
detection significance: $\nu_\text{e}$, $\nu_\text{size}$, and $\nu_\text{flux}$.  These can be
estimated using analytic formulae in some cases, or using measurements of multiple noise
realizations for each object.  One of our goals is to compare the behavior of these different
estimators of detection significance and relate them to $\nu_\text{ideal}$.

\subsection{Pixel noise models}\label{subsec:noisefields}

The noise fields used in this work are stationary (noise variance constant across the image)
and Gaussian-distributed.  This approximation is appropriate in the limit that the sky background
dominates and is large enough that a Poisson distribution is nearly Gaussian, which describes most
galaxies used for weak lensing measurements from the ground and a smaller but non-negligible
fraction from space.

Our uncorrelated noise fields are simply drawn from the appropriate Gaussian distribution, with a
new random number drawn for each pixel.  The correlated noise fields include the full
scale-dependent and direction-dependent correlation between noises in adjacent pixels in the images
used for weak lensing science in the COSMOS survey
\citep{2007ApJS..172..196K,2007ApJS..172....1S,2007ApJS..172...38S}.  These correlations largely
arise due to choices in how the image resampling and combination process is done.  Figure 2 in
\cite{2015A&C....10..121R} illustrates the noise correlation function for these noise fields; as
shown, when moving one pixel in the horizontal or vertical direction, this correlation function is
$\sim 0.4$ to $0.5$, while moving two pixels away results in it decreasing by another factor of 10.

This specific noise correlation function is adopted as a
practical example, but we caution that the details of our results (for example, detection
significances without vs.\ with correlated noise) depend on the adopted level of
correlations, which will in general differ for different datasets and analysis methods.  We explore variations in our default level of noise correlation in
Section~\ref{subsec:imageprop} and find that our overall conclusions are valid for
significantly stronger and weaker noise correlations, as well.  Thus, while coadded ground-based data will
typically have weaker correlation patterns than our default (since the images are typically
resampled to a pixel scale comparable to the native one, rather than a significantly smaller one),
our overall conclusions should apply to typical ground-based data.

Finally, a subset of our simulations include anisotropies in the correlated noise.  Correlated noise
is in general anisotropic, 
due (for example) to direction-dependent corrections for charge transfer inefficiency and camera
distortion.  Our tests reveal 
the degree to which these low-level anisotropies in the correlated noise field can contaminate
per-galaxy shear estimates.

\section{Simulation and analysis methods} \label{sec:methods} 

Image generation (Sec.~\ref{subsec:simgeneration}) and the measurement process
(Sec.~\ref{subsec:measurement}) is performed in a single seamless pipeline using \galsim. Galaxy and
PSF profiles are convolved and rendered using Fourier-based rendering methods
\citep{2015A&C....10..121R}.  Many independent noise realizations (up to $10^5$) are generated for
each value of SNR, and analyzed using the shape measurement methods in Sec.~\ref{subsec:shapemeas}.

\subsection{Simulation generation}\label{subsec:simgeneration}
\subsubsection{Galaxy Profile}

For this work, galaxy light profiles are represented using the \sersic\ profile \citep{sersic63}, a
family of light profiles that can describe a wide range of galaxies.
The surface brightness of a \sersic\ profile varies as 
\begin{equation}
I(r) = \frac{F}{a(n) \refresp{r_{1/2}}^2} \exp{[-b(n)(r/\refresp{r_{1/2}})^{1/n}]} \label{eq:sersic},
\end{equation}
where $F$ is the total flux, \refresp{$r_{1/2}$} is the
half-light radius, and $a(n)$ and $b(n)$ are known functions with
numerical solutions. 
Flux normalization determines $a(n)$, while $b(n)$ is set by requiring that the 
half-light radius enclose half of the flux.

Our tests use sixty possible \sersic\ galaxy profiles with one of five values 
of \sersic\ index, two values for each shape component, and three values of size.
Each galaxy consists of a single \sersic\ component. 
Table~\ref{T:groundparams} lists the values used for these parameters. 
	\begin{table} 
		\begin{center}
			\begin{tabular}{|c || c |} 
				\hline $n$ & $1, 2, 3, 4, 5$ \\ \hline 
				$e_1$&$-0.45$, $0.45$\\ \hline 
				$e_2$ & $-0.45$, $0.45$\\ \hline 
				$r_{1/2}$&$0.4\arcsec, 0.7\arcsec, 1\arcsec$\\ \hline 
				PSF& Kolmogorov (FWHM: 0.7\arcsec)\\ \hline
				Pixel Scale & 0.2\arcsec / pixel\\ \hline
			\end{tabular} 
			\caption{
				\label{T:groundparams}Parameters used for our main set of simulated images from
                ground-based telescopes.}
		\end{center} 
	\end{table} 

\subsubsection{PSF and pixel response}

	The simulated PSF is a circular Kolmogorov profile with a full width at half maximum (FHWM)
	 of 0.7\arcsec.  This choice is representative of long-exposure times from ground-based
     telescopes at sites with good imaging conditions. 
Our simulated ground-based images have a pixel scale of $0.2$\arcsec\ per pixel.

\subsubsection{Image rendering} 

	The galaxy and PSF (including the Kolmogorov profile and pixel response) are convolved
    using Fourier-space rendering in \galsim, and the light profiles are sampled onto a $50 \times 50$ pixel image 
    using the chosen pixel scale.
 
\begin{figure} 
	\begin{center}
	\includegraphics[scale=1.1]
	{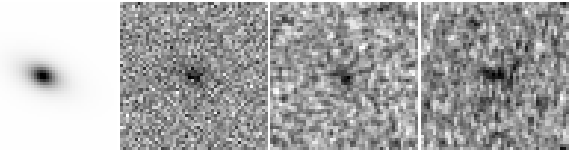}
	\caption{\label{F:sampleimage} Examples of postage stamp images 
	created by \galsim, for a single choice of \sersic\ profile parameters.  The color-scale is
    linear in the flux. From left to right, the panels show the image for $\nuu{ideal}=\infty$ (no
	noise added) and that same image with uncorrelated noise, 
	correlated noise, and correlated noise that was sheared with $(e_1=0.5,e_2=-0.5)$. All but the
    noise-free image have noise variance chosen such that 
	$\nuu{ideal}=15$ (one of the lowest SNR values used for this study). The case of sheared
    correlated noise is only used in Sec.~\ref{subsubsec:anisotropy}.}
	\end{center}
\end{figure} 

\subsubsection{Noise field}

For each value of $\nuu{ideal}$, Eq.~\eqref{eq:sndef} was used to calculate a noise variance to be
added to the initially noise-free postage stamp image.  
Many noise realizations with that variance are then generated, using 
uncorrelated and correlated noise as described in Sec.~\ref{subsec:noisefields}. 
Example images are shown in Fig.~\ref{F:sampleimage}, with the sheared correlated noise case only
used in Sec.~\ref{subsubsec:anisotropy}.

\subsection{Measurement process}\label{subsec:measurement}

Measurements of flux, size ($\sigma$), and shape ($e_1, e_2$) were
taken on as many as $10^5$ different realizations of correlated and
uncorrelated noise for the same galaxy image. In addition to quantifying
biases in the above quantities as a function of the idealized
detection significance \nuu{ideal}, the many noise realizations are also used to estimate the effective
detection significances defined in Sec.~\ref{sec:background}, namely
\nuu{e}, \nuu{size}, \nuu{flux}.  These are strictly below 
\nuu{ideal} (which assumes perfect information about the light
profile) and lower for correlated noise than for \refresp{un}correlated noise.

Flux, size, and shape measurements
from a single galaxy image were considered failures and discarded 
if the code raised a fatal error, or if one of the measurement of flux or size fell 
outside a generalized $5\sigma$ range centered at the mean.  
This range is determined by independently constructing the
probability distribution function (PDF) of the values of both log(size) and log(flux),
 and then estimating $\sigma_\text{low}$ and
$\sigma_\text{high}$ as the difference between the 50th and 16th
percentile, and the 84th and 50th percentile, respectively for each PDF.  Measured
values that are less than $5\sigma_\text{low}$ below the mean, or more
than $5\sigma_\text{high}$ above the mean, are excluded as failures.

\begin{figure} 
	\begin{center}
	\includegraphics[width=\columnwidth,angle=0]
	{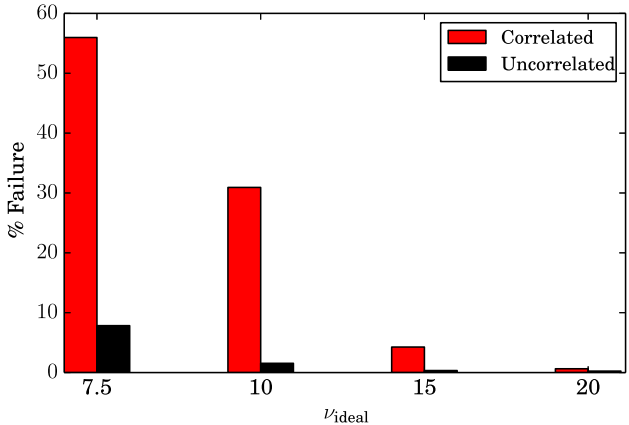}
	\caption{\label{F:fails} The percentage of measurement failures as a function of \nuu{ideal} is
      shown for the case of uncorrelated and correlated noise, for one particular galaxy model.  
	For $\nuu{ideal}> 20$ (not shown) the failure rate falls consistently below 1 per
    cent.}
	\end{center}
\end{figure}

Any dataset consisting of the measurements of flux, size, and shape for all noise realizations of a
particular galaxy model and noise variance with a failure rate above 20 per cent is discarded, out of
a concern that the remaining 80 per cent of successful measurements
might be a non-representative sample.  Failure rates this high only
occur for the case of correlated noise with $\nuu{ideal}=7.5$ and $10$; the failures arise primarily
due to a failure of the adaptive moment routine to converge in a reasonable number of iterations at
low effective SNR.
As a result, there are two fewer points for correlated noise results
than for uncorrelated noise on every plot in
Section~\ref{sec:results}.  Failure rates as a function of \nuu{ideal}
are shown for one particular galaxy model in Fig.~\ref{F:fails}, which
shows a rapid convergence to a very low failure rate at
$\nuu{ideal}=20$ for both correlated and uncorrelated noise. 
\refresp{In principle, a dependence of the failure rate on galaxy position angle and/or shape could
  result in a form of shear selection bias that depends on the type of noise (correlated vs.\
  uncorrelated).  We will explore this possibility in Sec.~\ref{subsubsec:shapedep}.}

\section{Results} \label{sec:results}

Our results are organized as follows. 
Sec.~\ref{subsec:biasideal} shows the biases in galaxy shapes for a 
single galaxy model as a function of the ideal detection significance
\nuu{ideal} for correlated and uncorrelated noise.  
Sec.~\ref{subsec:nu-eff} shows our results for the effective
detection significance defined in various ways, 
 again for both correlated and uncorrelated noise.

An obvious question to ask is that since uncorrelated and correlated
noise with different variance (zero-lag value of the noise correlation
function) results in different effective detection significance but
the same \nuu{ideal}, are the shear biases determined in
Sec.~\ref{subsec:biasideal} better quantified in terms of the
effective detection significance?  And, if defined that way, do the
results with uncorrelated and correlated noise follow identical
trends?  We address this point in Sec.~\ref{subsec:remapping}.  The
results up to that point are for a specific galaxy model and to a
noise correlation structure similar to that seen in COSMOS weak
lensing images.  The dependence of these results
on galaxy properties such as size, \sersic\ index, and shape
(Sec.~\ref{subsec:galprop}), and on PSF and noise field properties
(Sec.~\ref{subsec:imageprop}) are shown next.  Finally, the extension of our results to another
galaxy shape measurement code,
\code{im3shape}, is shown in Sec.~\ref{subsec:generalizing}.  

\subsection{Shape bias}\label{subsec:biasideal}

Unless otherwise specified-- i.e., in Sec.~\ref{subsec:galprop} 
and~\ref{subsec:imageprop} -- the plots and data presented in this and 
subsequent subsections are for 
a galaxy with $(e_1,e_2)=(0.45,-0.45)$, $r_{1/2}=0.7\arcsec$, and $n=2$. 
These midrange values were chosen to represent the overall conclusions drawn 
from analysis of all combinations of galaxy parameters in Table~\ref{T:groundparams}.

\begin{figure} 
	\begin{center} 
	\includegraphics[width=\columnwidth]{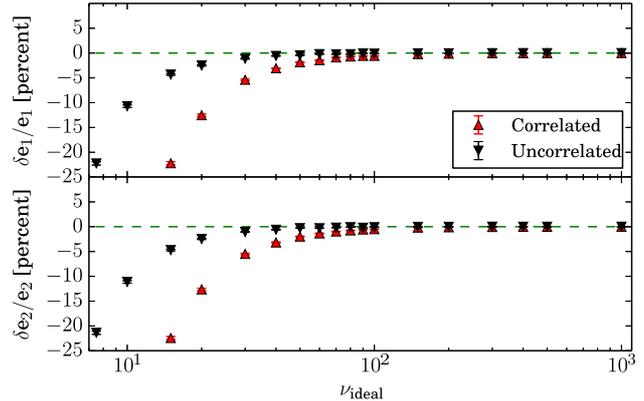}
	\caption{\label{F:shape} Shape measurements as a function of the
      ideal detection significance $\nu_\text{ideal}$, for
	images with correlated and uncorrelated noise. Results for the two
    shape components are shown in separate panels.  $\delta e_i/e_i$
	is the relative bias in shape component $i$, plotted here as a 
	percentage.  Points are shown with errorbars, which 
    are typically smaller than the points themselves.}
	\end{center} 
\end{figure}

Fig.~\ref{F:shape} shows the fractional bias in galaxy shape measurements\footnote{For a weak
  lensing measurement, what actually matters is the bias in the ensemble average {\em shear}
  rather than the per-galaxy {\em shapes}.  The bias in that ensemble shear estimate is a weighted
  average bias of the per-galaxy shape biases, which will in general depend on SNR, resolution,
  morphology, and other quantities.  Nonetheless, for this study 
we focus on galaxy shape biases, as a simple way to gain some initial insight into the impact of
correlated noise.} as 
a function of the ideal detection significance, \nuu{ideal}, for this galaxy model. The
results are quite similar for both components, so future plots of
shear biases are restricted to a single component.  There
is clear convergence to a bias of zero at $\nuu{ideal}\approx 100$ for
both the correlated and uncorrelated cases.  At lower detection
significance, the bias becomes negative, indicating an overly round
shape.  The sign of this effect is consistent with simulation-based
investigations of the shear biases of the re-Gaussianization method
\citep{2012MNRAS.420.1518M,2015MNRAS.450.2963M}.  

At a fixed value of
\nuu{ideal}, the bias is stronger for correlated noise than for
uncorrelated noise, but the curves have approximately the same shape.
This trend suggests an interpretation in terms of some effective,
empirically-determined detection significance, an idea that will be
explored further in Sec.~\ref{subsec:remapping}.

\subsection{Effective detection significance}\label{subsec:nu-eff}

This section contains empirical estimates of the effective
detection significance \nuu{eff} (\nuu{flux}, \nuu{size}, and
\nuu{e}), using the same galaxy model as in the previous subsection. 
The results shown here use $10^5$
noise realizations in order to robustly measure uncertainties in
measured quantities.

\begin{figure*} 
	\begin{center} 
	\includegraphics[width=6in]{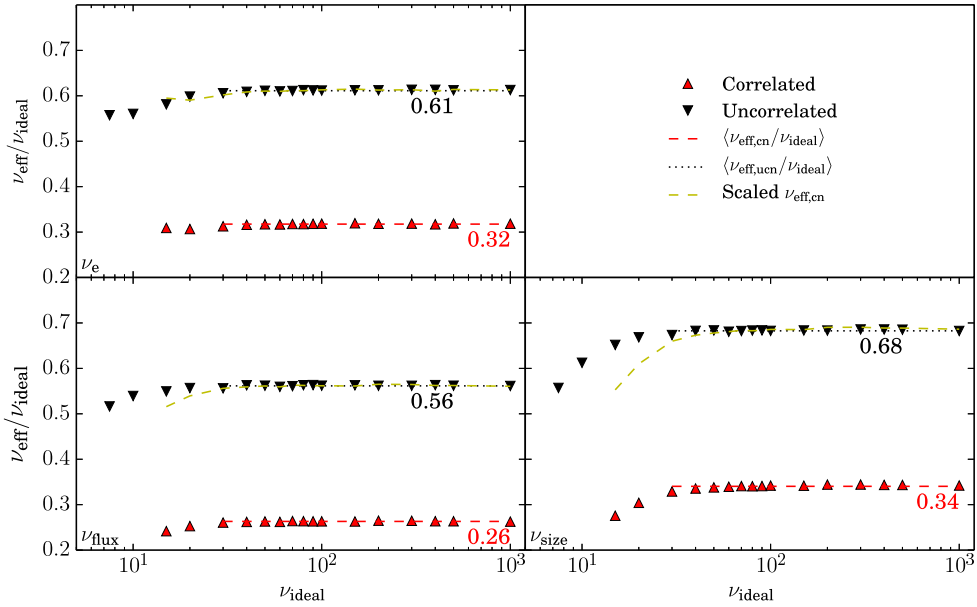}
	\caption{\label{F:errmap}Three different measures of the 
	effective detection significance \nuu{eff} (top left: \nuu{e};
    lower left: \nuu{flux}; lower right: \nuu{size}) are shown as $\nuu{eff}/\nuu{ideal}$ as a
    function of \nuu{ideal}.  The 
    average values of $\nuu{eff}/\nuu{ideal}$ across all \nuu{ideal}
    values is shown as $\langle\nuu{eff}/\nuu{ideal}\rangle$ (horizontal lines, with the
    average value given directly below each line).  Only $\nu_{e_1}$ is 
	plotted as \nuu{e}, since it is identical to results for the other
    shape component. 
	A ``Scaled	\nuu{eff,ucn}'' line was constructed by multiplying the 
	red points (correlated noise) by the ratio of the
    $\langle\nuu{eff}/\nuu{ideal}\rangle$ lines for uncorrelated vs.\
    correlated noise. This allows for an easier comparison between the
  shapes of the curves for correlated and uncorrelated noise.}
	\end{center}
\end{figure*}
Measurements for a range of noise variance values spanning over two
orders of magnitude ($7.5\le \nuu{ideal}\le 1000$), shown in
Fig.~\ref{F:errmap}, reveal that for the majority of that range, the
ratio of effective to ideal detection significance
$\nuu{eff}/\nuu{ideal}$ is consistent with a single constant value
that is significantly below $1$, as in \cite{2015MNRAS.450.2963M}.
That ratio takes on different values in the case of uncorrelated noise
and correlated noise, with the former being higher (as expected) by
roughly a factor of 2.  For $\nuu{ideal}\lesssim 20$, this simple
$\nuu{eff}\propto\nuu{ideal}$ relation is violated, with $\nuu{eff}$
falling below the linear relation.

For fixed $\nuu{ideal}>20$, the \nuu{eff} values satisfy $\nuu{flux}<\nuu{e}<\nuu{size}$.  This
hierarchy is not necessarily consistent for galaxy models with a different size or shape, or for
different correlated noise fields, as will be shown in subsequent sections.

\subsection{Remapping shape biases with \nuu{eff}} \label{subsec:remapping}

The results from Sec.~\ref{subsec:nu-eff} can be used to reinterpret
those from Sec.~\ref{subsec:biasideal} in terms of effective detection
significances \nuu{eff}. Starting with the results in Fig.~\ref{F:errmap}, we
can scale the values of \nuu{ideal} for each dataset  by
$\langle\nuu{eff}/\nuu{ideal}\rangle$ to plot galaxy shape biases as a function of
\nuu{eff}. \refresp{This ratio, $\langle\nuu{eff}/\nuu{ideal}\rangle$, is taken to be a single
  number that can be read off from Fig.~\ref{F:errmap} for each type of detection significance
  and noise\footnote{One might ask why we do not simply
  use the measured values of $\nuu{eff}$ directly, instead of applying
  the simple linear scaling $\nuu{eff}\propto\nuu{ideal}$.  We adopt
  this simpler procedure because in a study aimed at calibrating shear
  biases, it seems more likely that one would determine roughly how
  the detection significance scales with the square root of the noise
  variance, and apply an overall correction factor, just as we are
  doing here.}.} The
key step is to do this rescaling separately for uncorrelated and correlated noise.  Our goal is to
determine whether, after doing so, the highly offset curves in
Fig.~\ref{F:shape} for correlated vs.\ uncorrelated noise lie on top
of each other.  We
focus on the region where the shape bias is significantly nonzero,
$\nuu{ideal}\le 40$, since that is the regime where noise bias 
must be calibrated using simulations.

\begin{figure} 
	\begin{center} 
	\includegraphics[width=\columnwidth]{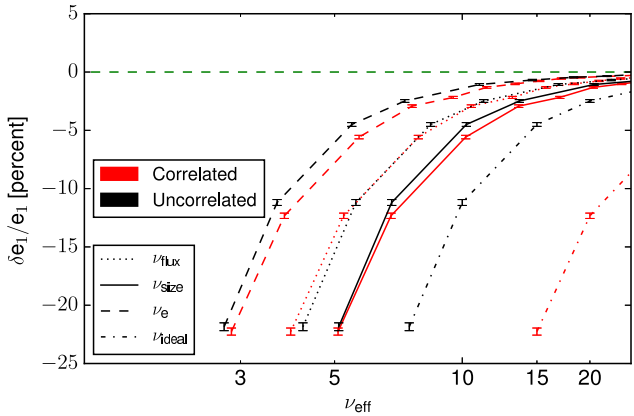}
	\caption{\label{F:6shape} Biases in shape measurements as a function of
      $\nu_\text{eff}$, for different choices of \nuu{eff} (\nuu{e},
      \nuu{flux}, and \nuu{size}), for both uncorrelated and correlated
      noise.  For comparison, the results from Fig.~\ref{F:shape}
      as a function of \nuu{ideal}
      are also shown.  Thus, the plot shows just two curves (correlated and uncorrelated noise biases),
      each repeated at four distinct sets of positions along the $x$ axis, with horizontal
      rescalings determined from Fig.~\ref{F:errmap}.  Results were determined using $10^5$ noise
      realizations.  Note the more limited horizontal axis range compared to
      Fig.~\ref{F:shape}, chosen to accentuate the region of
      the curves with a nonzero bias.
}
	\end{center} 
\end{figure}
Fig.~\ref{F:6shape} shows the relative bias in one shape component 
as a function of the three \nuu{eff} values and \nuu{ideal}, for both
uncorrelated and correlated noise.  In Fig.~\ref{F:shape}, differences
in the shear calibration bias for uncorrelated noise vs.\ correlated
noise at fixed \nuu{ideal} were typically factors of $\sim 6$.  It is
clear that the use of any of our \nuu{eff} options in
Fig.~\ref{F:6shape} significantly reduces this difference to at most
20 per cent.  However, of the three options, \nuu{flux} 
 most reduces the difference between the shear biases with
correlated and uncorrelated noise.  When plotting the bias in galaxy shape 
as a function of \nuu{flux}, the curves for
uncorrelated and correlated noise lie nearly on top of each other.
\nuu{size} is the next best option, and \nuu{e} is, interestingly, the
worst option of the three.

The results of this subsection suggest that when applying
simulation-based noise bias corrections to galaxy shape estimates,
 results from noise correlation-free
simulations could be used to correct the results in data that includes
noise correlations, as long as the impact of noise correlations on 
the detection significance are accounted for. \refresp{To quantify what error might be induced by
  doing so, we take the points in Fig.~\ref{F:6shape} for
  uncorrelated noise, and find a three-parameter fitting function that describes the bias 
  as a function of \nuu{eff,ucn}, using \nuu{flux} to define the effective detection significance.  This function is}
$$
\refresp{\text{Bias (\%)} = -0.074 - \frac{581.1}{\nuu{eff,ucn}^{2.29}}}.
$$
\refresp{In reality the expansion is more
likely a constant plus $1/\nuu{eff,ucn}^2$ plus higher order terms as shown in
\cite{2004MNRAS.353..529H}, but we used a single term with varying power-law index to effectively
include some higher order behavior without introducing many degrees of freedom into the fit. 
We evaluate this function at the values of \nuu{eff,cn} in Fig.~\ref{F:6shape}, and calculate
the residual bias for the correlated noise case, defining the residual as the actual bias on
Fig.~\ref{F:6shape} minus the bias from the fit.  The RMS value of the
residual bias over all points on the curve is 
0.3\%, more than an order of magnitude below the typical biases for $\nuu{eff} <20$.}

\subsection{Dependence on galaxy properties}\label{subsec:galprop}

This section illustrates the dependence of our results on the
properties of the simulated galaxies, going beyond the single
galaxy model used in Sec.~\ref{subsec:biasideal}.  We test how the
ratio $\langle\nuu{eff,ucn}\rangle/\langle\nuu{eff,cn}\rangle$ inferred from
Fig.~\ref{F:errmap}, and the efficacy of the remapping described in
Sec.~\ref{subsec:remapping}, varies with different properties
described below.

\subsubsection{\sersic\ index and size}

The same process as in previous sections was carried out for fifteen different
combinations of parameters (five values of \sersic\ index, three
values of $r_{1/2}$ and a single value of $(e_1,e_2)=(0.45,-0.45)$)
 originally listed in Table~\ref{T:groundparams}.

Carrying out the exercise in Fig.~\ref{F:errmap} to find the
relationship between $\nuu{eff}/\nuu{ideal}$ for various choices of
\nuu{eff} reveals variations in those ratios that are
as large as \refresp{4}0 per cent across the entire range in \sersic\ index or
$r_{1/2}$.  For a fixed value of \nuu{ideal} (i.e., noise variance) and
galaxy flux, \nuu{eff} tends to be largest for small galaxies (low
$r_{1/2}$) and low \sersic\ index.  This makes sense, since the
light profile covers fewer pixels and thus has effectively less noise in that case. 

%
%

\begin{figure*} 
	\begin{center} 
	\includegraphics[width=6in]{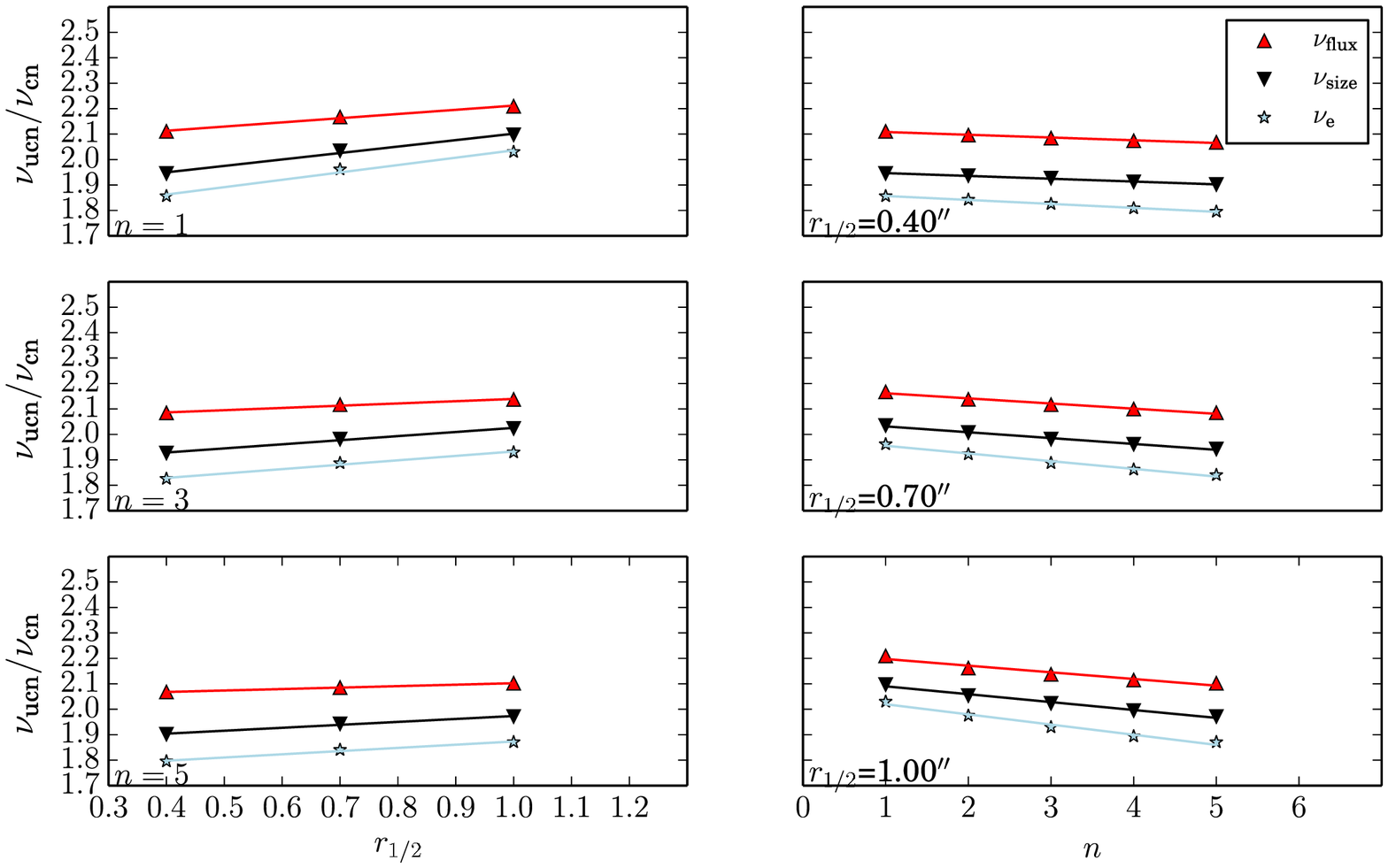}
	\caption{\label{F:remappings} The ratio of detection
      significance for uncorrelated vs.\ correlated noise,
      $\langle\nuu{eff,ucn}\rangle/\langle\nuu{eff,cn}\rangle$
      (labeled more simply as $\nuu{ucn}/\nuu{cn}$), as a
      function of galaxy properties.  The left (right) column show
      trends with galaxy size, $r_{1/2}$, at fixed \sersic\ $n$ (with
      \sersic\ $n$ at fixed $r_{1/2}$).  The equations for the plotted trendlines
	can be found in Table~\ref{T:remappings}.}

	\end{center} 
\end{figure*}

Tests of the impact of correlated vs.\ uncorrelated
noise use the ratio 
$\langle\nuu{eff,ucn}\rangle/\langle\nuu{eff,cn}\rangle$, as shown in
Fig.~\ref{F:remappings}.  The left column shows that at fixed
\sersic\ index, this ratio is a weakly increasing function of
$r_{1/2}$, for all three types of \nuu{eff}.  The strength of the
trend is larger at low fixed $n$, but still less than 10 per cent in
the most extreme case (and therefore weaker than the individual trends
in \nuu{eff,ucn} or \nuu{eff,cn} with $r_{1/2}$).  The right column shows that at fixed size,
this ratio is a weakly decreasing function of \sersic\ $n$, again
reaching a maximum variation of 10 per cent across the entire range of
galaxy parameters and being strongest for the largest galaxies
considered.  

This stability in
$\langle\nuu{eff,ucn}\rangle/\langle\nuu{eff,cn}\rangle$ is fairly
remarkable given that the range of galaxy sizes is a factor of 2.5
and the range of \sersic\ $n$ goes from $1$ to $5$. \refresp{The stability derives from the fact
  that $\nuu{eff,ucn}/\nuu{ideal}$ and $\nuu{eff,cn}/\nuu{ideal}$ both have stronger trends with
  galaxy properties as mentioned above, but the trends are sufficiently similar that they largely cancel out in the
  ratio $\nuu{eff,ucn}/\nuu{eff,cn}$.  Any scheme that aims to correct for calibration
biases in shape measurements using a combination of simulations with uncorrelated and correlated
noise would need to account properly for the differences between $\nuu{eff,ucn}/\nuu{ideal}$ and $\nuu{eff,cn}/\nuu{ideal}$.}
 
\begin{figure} 
	\begin{center} 
	\includegraphics[width=\columnwidth]{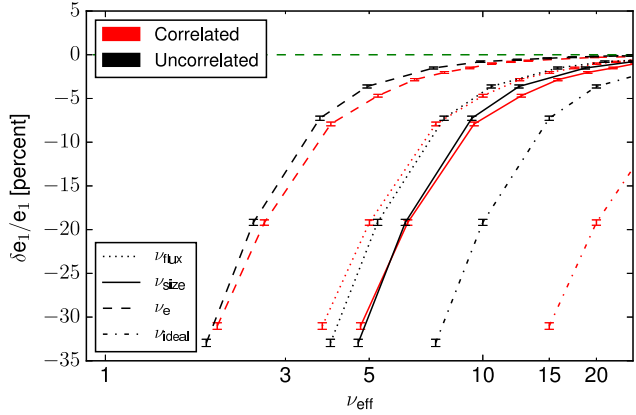}
	\caption{\label{F:extreme_6shape}
Biases in shape measurements as a function of
      $\nu$, for different choices of \nuu{eff} (\nuu{e},
      \nuu{flux}, and \nuu{size}) and \nuu{ideal}, for both uncorrelated and correlated
      noise.  Unlike Fig.~\ref{F:6shape}, this plot uses a different galaxy model with $n=5$ and 
	$r_{1/2}=1$\arcsec.  The remapping by \nuu{eff} is still successful in 
	reducing the horizontal offset in the galaxy shape bias curves for correlated and uncorrelated 
	noise. 
}
	\end{center} 
\end{figure}

The parameters of the lines shown in Fig.~\ref{F:remappings} can be
found in Table~\ref{T:remappings}.  We also confirm that the success
in remapping shape bias vs.\ \nuu{eff} to match up the uncorrelated
and correlated noise results (Sec.~\ref{subsec:remapping}) carries over to all other values of $n$ and
$r_{1/2}$ tested in this section.   A single example is shown in
Fig.~\ref{F:extreme_6shape}.

\begin{table*} 
		\begin{center}
			\makebox[\textwidth][c]{\begin{tabular}{| c || c |} 
\hline 
$n=1$ & $r_{1/2}$= 0.4\arcsec\\ \hline
	\begin{tabular}{ ccc }
	Flux & Size & Shape \\ \hline
	\rmxb{0.16}{2.16} & \rmxb{0.25}{2.03} & \rmxb{0.29}{1.95}
	\end{tabular} & 
	\begin{tabular}{ ccc }
	Flux & Size& Shape \\ \hline
	\nmxb{-0.01}{2.09} & \nmxb{-0.01}{1.92} & \nmxb{-0.02}{1.83}
	\end{tabular} \\
 \hline \hline
$n=3$ & $r_{1/2}$= 0.7\arcsec\\ \hline
	\begin{tabular}{ ccc }
	Flux & Size & Shape \\ \hline
	\rmxb{0.09}{2.11} & \rmxb{0.16}{1.98} & \rmxb{0.17}{1.88}
	\end{tabular} & 
	\begin{tabular}{ ccc }
	Flux & Size& Shape \\ \hline
	\nmxb{-0.02}{2.12} & \nmxb{-0.02}{1.99} & \nmxb{-0.03}{1.89}
	\end{tabular} \\
\hline \hline
$n=5$ & $r_{1/2}$= 1.0\arcsec\\ \hline
	\begin{tabular}{ ccc }
	Flux & Size & Shape \\ \hline
	\rmxb{0.06}{2.09} & \rmxb{0.12}{1.94} & \rmxb{0.13}{1.84}
	\end{tabular} & 
	\begin{tabular}{ ccc }
	Flux & Size& Shape \\ \hline
	\nmxb{-0.03}{2.15} & \nmxb{-0.03}{2.03} & \nmxb{-0.04}{1.94}
	\end{tabular} \\
\hline 
\end{tabular}}
			\caption{\label{T:remappings}
			The equations of the trendlines shown in Fig.~\ref{F:remappings},
			organized according to their positions in that plot. Here, 
			 $y$ represents $\nu_\text{ucn}/\nu_\text{cn}$ while $n'$ and $r_{1/2}'$ 
			 represent the varying quantities respectively, offset such that $n'=(n-2)$
			 and $r_{1/2}'=(r_{1/2}-0.7\arcsec)$. 
}
		\end{center}
\end{table*}	

\subsubsection{Shape Dependence}\label{subsubsec:shapedep}

The next test was for the dependence of these results on the galaxy shape using
galaxies with \sersic\ $n=2$ and $r_{1/2}=0.7\arcsec$, for 
$e_1={0.1, 0.3, 0.45}$ and $e_2=0$.  Since our results in prior
sections were very similar for the two shape components, only
$e_1$ was varied in this section.

\begin{figure} 
	\begin{center}
	\includegraphics[width=\columnwidth,angle=0]
	{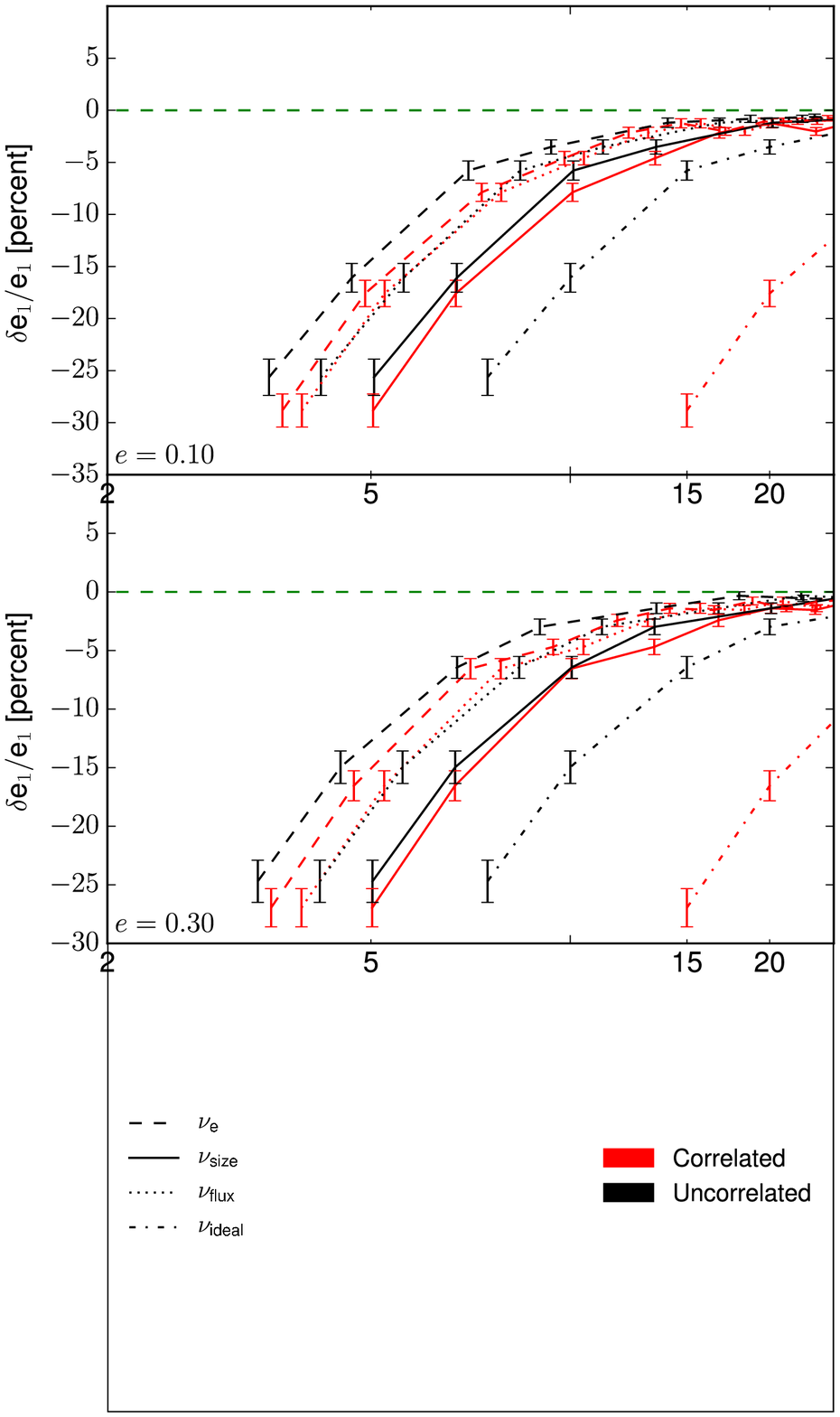}
	\caption{\label{F:shapes_shape} 
Biases in shape measurements as a function of
      $\nu$, for different choices of \nuu{eff} (\nuu{e},
      \nuu{flux}, and \nuu{size}) and \nuu{ideal}, for both uncorrelated and correlated
      noise.  Unlike Fig.~\ref{F:6shape}, this plot shows results for two other values of
      galaxy shape, for which the remapping by \nuu{eff} is successful in 
	reducing the offset in the galaxy shape bias curves for correlated and uncorrelated 
	noise. 
}
	\end{center}
\end{figure}

The ratio $\nuu{ucn}/\nuu{cn}$ 
varies very little ($<1$ per cent) over the range of shapes considered here.  Moreover,
Fig.~\ref{F:shapes_shape} demonstrates that the remapping successfully reduces the differences
between the shape bias for correlated and uncorrelated noise for the three values of $e_1$
considered.  This result demonstrates that our
results for $e_1=0.45$ can be considered as a general result,
independent of shape.

\refresp{As noted in Sec.~\ref{subsec:measurement}, selection effects that correlate with shape can
  cause a bias in ensemble shear estimation.  If this bias differs for correlated and uncorrelated
  noise, then that would also be a potential issue with remapping shear biases for the two noise
  types.  However, for the simulated galaxy and PSF profiles in this work, and the range of $|e|$
  explored in this section, we find that the failure rate is totally independent of $|e|$ for both
  correlated and uncorrelated noise.  It is also independent of the galaxy position angle for fixed
  $|e|$.  Note that in a more complex situation with anisotropic PSFs, it is possible that this
  would no longer be the case.}

\subsection{Dependence on image properties}\label{subsec:imageprop}

In this section, we explore the dependence of our results on the
imaging properties, including  the nature of the noise
correlations and the type of PSF.

\subsubsection{Different scale-length of noise correlations}\label{subsubsec:imagepropcor}

Our first tests are of the dependence of the results on the
scale-length of the noise correlations.  While our original 
noise correlation function corresponded to that in the COSMOS weak lensing science
images, \galsim\ permits users to ``magnify'' the noise correlation
function.  This operation corresponds to preserving the variance (zero-lag
correlation function) while changing the scale length of correlations.
Our tests involve linearly expanding the scale-length
of correlations by factors of $f_\text{exp}={0.5, 2.0}$, for a galaxy with
\sersic\ $n=2$, shape $(e_1,e_2)=(0.45,-0.45)$, and $r_{1/2}$=0.7\arcsec. 

\begin{figure} 
	\begin{center}
	\includegraphics[width=\columnwidth,angle=0]
	{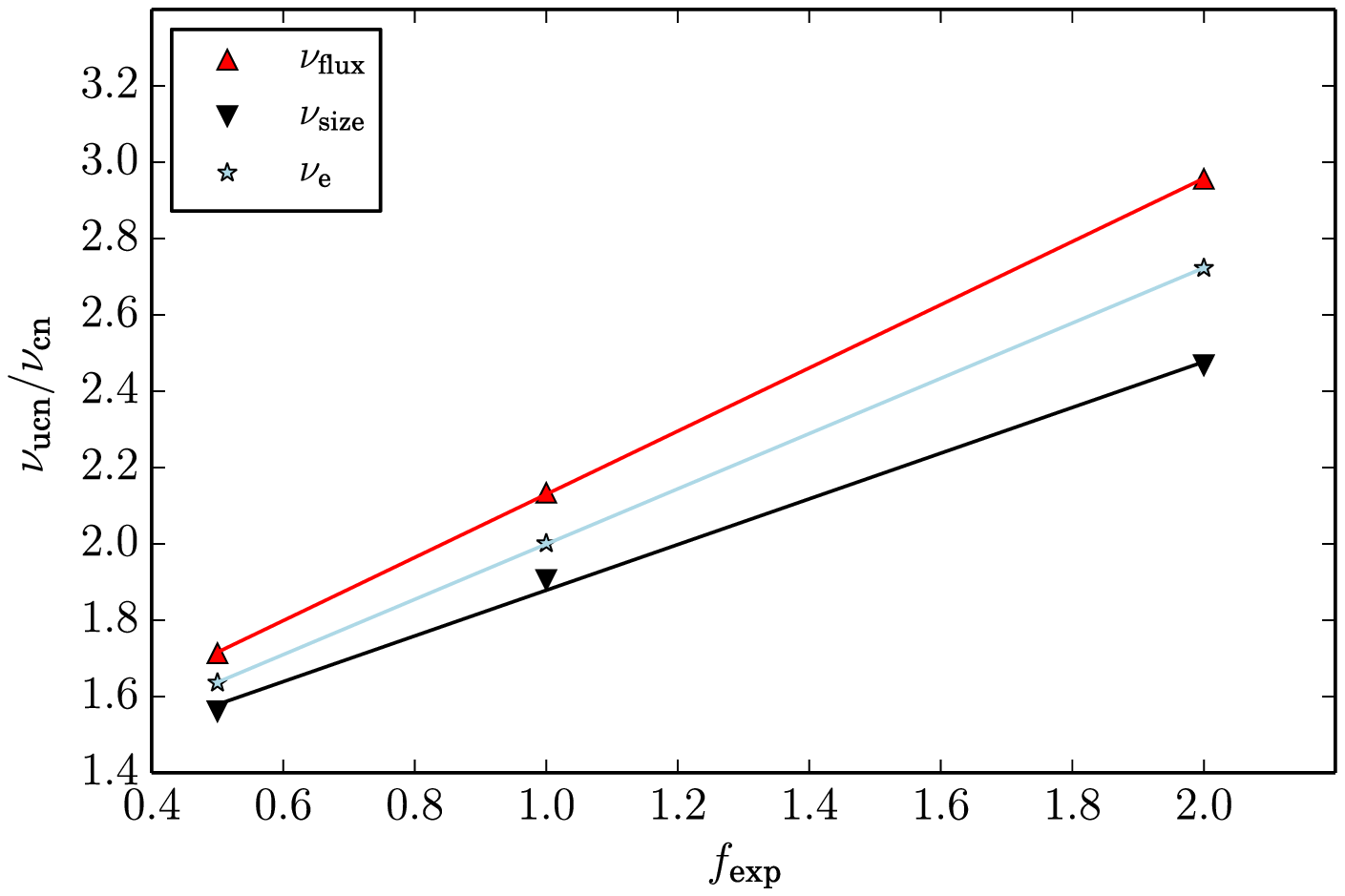}
	\caption{\label{F:coeffs_mag} The ratio $\nuu{eff,ucn}/\nuu{eff,cn}$
is shown as a function of $f_\text{exp}$, the linear expansion
factor that was applied to the scale length of the correlated noise
field.  The lines are the result of a linear fit to the points.}
	\end{center}
\end{figure}

The results for $\nuu{eff,ucn}/\nuu{eff,cn}$ are shown in
Fig.~\ref{F:coeffs_mag}.  Since only the correlated
noise field is being modified, these should be interpreted as modifications in
\nuu{eff,cn} with \nuu{eff,ucn} remaining fixed.  The ratio $\nuu{eff,ucn}/\nuu{eff,cn}$
increases linearly with $f_\text{exp}$, implying that the detection significance with
correlated noise, \nuu{eff,cn}, is smaller for large
$f_\text{exp}$.  The sign of this trend makes sense: for a fixed point variance, larger
$f_\text{exp}$ implies that the noise correlations extend over larger spatial scales, which should
decrease the effective detection significance for extended objects.

\begin{figure} 
	\begin{center}
	\includegraphics[width=\columnwidth,angle=0]
	{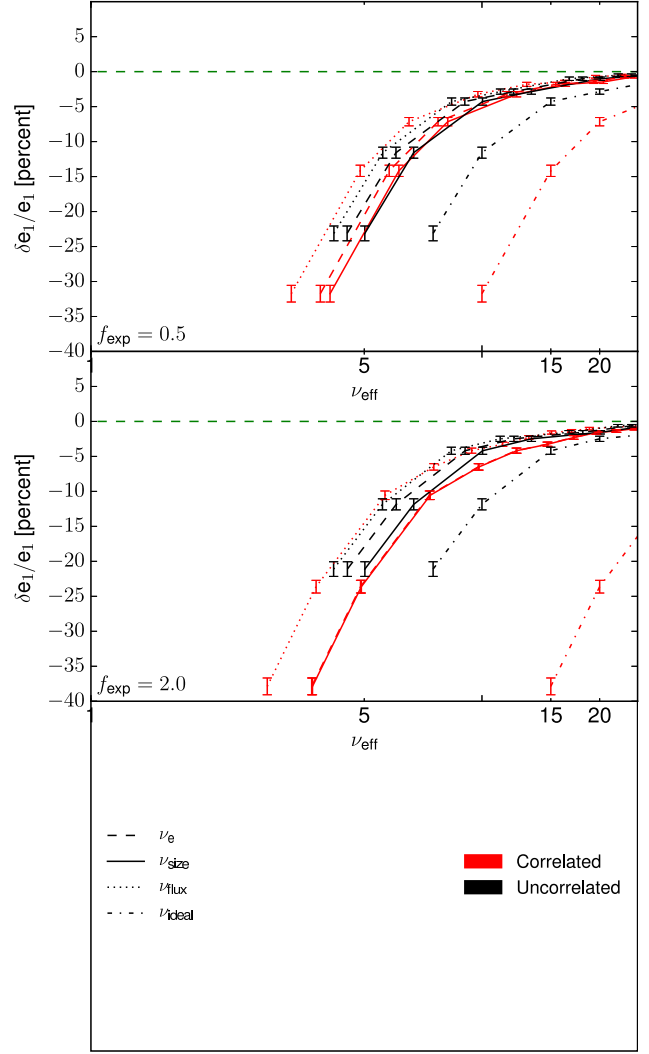}
	\caption{\label{F:mags_shape}
Biases in shape measurements as a function of
      $\nu$, for different choices of \nuu{eff} (\nuu{e},
      \nuu{flux}, and \nuu{size}) and \nuu{ideal}, for both uncorrelated and correlated
      noise.  Unlike Fig.~\ref{F:6shape} (which has $f_\text{exp}=1$, i.e., no modification in the
      correlation length of the noise correlations), this plot shows results for two different values of
      $f_\text{exp}$, a linear expansion of the scale length of noise correlations, for which the remapping by \nuu{eff} is successful in 
	reducing the offset in the galaxy shape bias curves for correlated and uncorrelated 
	noise. }
	\end{center}
\end{figure}

Fig.~\ref{F:mags_shape} shows that remapping the shape biases based on the $\nuu{eff}/\nuu{ideal}$
successfully matches up the shape biases for uncorrelated and correlated noise for all three values
of $f_\text{exp}$.  Comparison of the
panels indicates that this is a non-trivial statement, since the difference between the results for
uncorrelated and correlated noise as a function of the point variance can be quite large for large
values of $f_\text{exp}$.  Hence our results seem fairly robust to the scale-length of the noise
correlations.

It is possible that if the scale-length of the correlations became comparable to the galaxy size
itself, then additional effects would come into play.  The cases considered here are not in
this regime (which should not be relevant for too many galaxies that will be used for weak lensing),
and we defer this question to future work.

\subsubsection{Anisotropy in noise correlations}\label{subsubsec:anisotropy}

The correlated noise field  used throughout this work has a mild level of anisotropy in
the noise correlations (a few per cent).  Generally, unless explicitly
accounted for\footnote{Most shear estimation routines do not have a
  means to account for correlated noise fields, regardless of their
  anisotropy level.} these anisotropies are expected to be imprinted
at some level in the measurements of galaxy shape, particularly at low
detection significance.  This effect is explored in this section, with a
galaxy with \sersic\ $n=2$, $r_{1/2}$=0.7\arcsec, and $(e_1,e_2)=(0.45,-0.45)$, using our original
correlated noise field as well as versions that have an additional shear applied to the noise correlation
function.  To make the trends visually apparent, the noise correlations are sheared by an
exaggerated amount, 
$e_\text{noise}=0.5$ (see right-most panel of 
Fig.~\ref{F:sampleimage} for an example image).

In this case, what matters are the results with shear applied to
the correlated noise compared to the results with the original correlated noise.  More specifically, when measuring
$e_{i,\text{orig}}$ for shape component $i$ and the original noise field, and $e_{i,\text{sh}}$ with
the sheared noise field, for the case where the expected shape is $e_{i,\text{exp}}$, the following
differences are calculated and plotted:
	\begin{align*}
	\Delta\delta\text{e}_{i}/\text{e}_i&=\frac{\text{e}_{i,\text{sh}}-\text{e}_{i,\text{exp}}}{\text{e}_{i,\text{exp}}}
	-\frac{\text{e}_{i,\text{orig}} - \text{e}_{i,\text{exp}}}{\text{e}_{i,\text{exp}}}\\
	&=\frac{\text{e}_{i,\text{sh}}-\text{e}_{i,\text{orig}}}{\text{e}_{i,\text{exp}}}. 
	\end{align*}
\begin{figure} 
	\begin{center}
	\includegraphics[width=\columnwidth,angle=0]
	{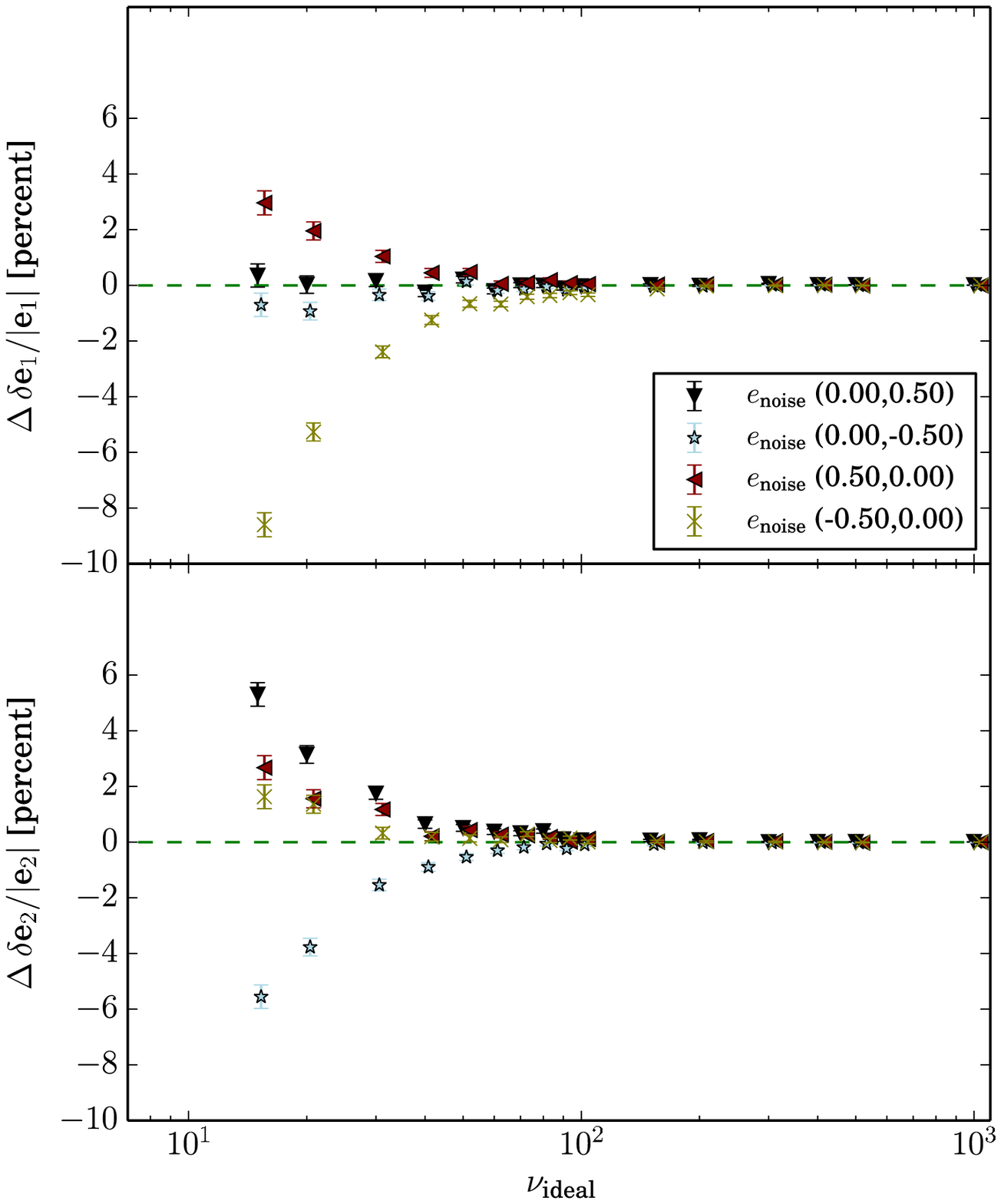}
	\caption{\label{F:shshape} Modifications to the galaxy shape bias when shearing the correlated noise field by the
      amounts given in the legend are shown as a function of the ideal detection significance,
      \nuu{ideal}.}
	\end{center}
\end{figure} 

The results are shown in Fig.~\ref{F:shshape}.  When shearing the correlated noise in
one component, the dominant effects on the shape bias are in that component, but they are not
completely zero for the other component.  The sign of the effect is as expected, with positive noise
shear in one component translating into a positive shape bias.  For the lowest \nuu{ideal} value
considered ($\nuu{ideal}=15$), the magnitude of the additional shape bias is typically of order 5 per cent.  We
expect this to scale linearly with the noise shear, so for more realistic values like $e_\text{noise}=0.1$,
the maximum shape bias would be of order 1 per cent for the lowest SNR galaxies in our simulated
sample.  This effect is therefore large enough to matter for upcoming surveys
that require understanding of shear calibration biases at the sub-percent level.

An additional source of bias not considered here is selection bias.  If there is anisotropic
correlated noise, and if galaxies are more or less likely to be selected for measurement if their
shape aligns in some particular way with respect to that noise field, then this will cause an
additional bias in ensemble shear estimates that is not captured by our tests.

\subsubsection{Ground-like versus space-like data}

To test the impact of the PSF type and sampling, 
simulated space-based data was also generated, with parameters in Table~\ref{T:spaceparams}, and
used for the same analysis as was already presented for the ground-like simulations.  For a galaxy
model with $(e_1,e_2)=(0.45,-0.45)$, $r_{1/2}=0.2\arcsec$, and \sersic\ $n=2$, 
the ratio of $\nuu{eff}/\nuu{ideal}$ is qualitatively similar to that in
Fig.~\ref{F:errmap}, in the sense that there are flat curves for $\nuu{ideal}\gtrsim 20$.  The
actual ratios differ, with values of $\nuu{e}/\nuu{ideal}=0.83$ and $0.42$ for uncorrelated and
correlated noise, respectively; $\nuu{flux}/\nuu{ideal}=0.43$ and $0.20$; and
$\nuu{size}/\nuu{ideal}=0.47$ and $0.23$.  Thus, in all three cases, the ratio of $\nuu{eff}$ for
correlated vs.\ for uncorrelated noise is $\sim 2$.  However, the hierarchy from
Fig.~\ref{F:errmap} ($\nuu{flux}<\nuu{e}<\nuu{size}$) is not respected here, implying that the
hierarchy depends on the details of the type of data.  Note that the galaxy model here is smaller
than for ground-like data, with $r_{1/2}=0.2\arcsec$ instead of $0.7\arcsec$.  This choice, which is
meant to reflect the fact that the typical galaxy that is analyzed in space-based datasets is
smaller than can be easily resolved in ground-based datasets, may in part be
responsible for the change.

\begin{figure} 
	\begin{center}
	\includegraphics[width=\columnwidth]
	{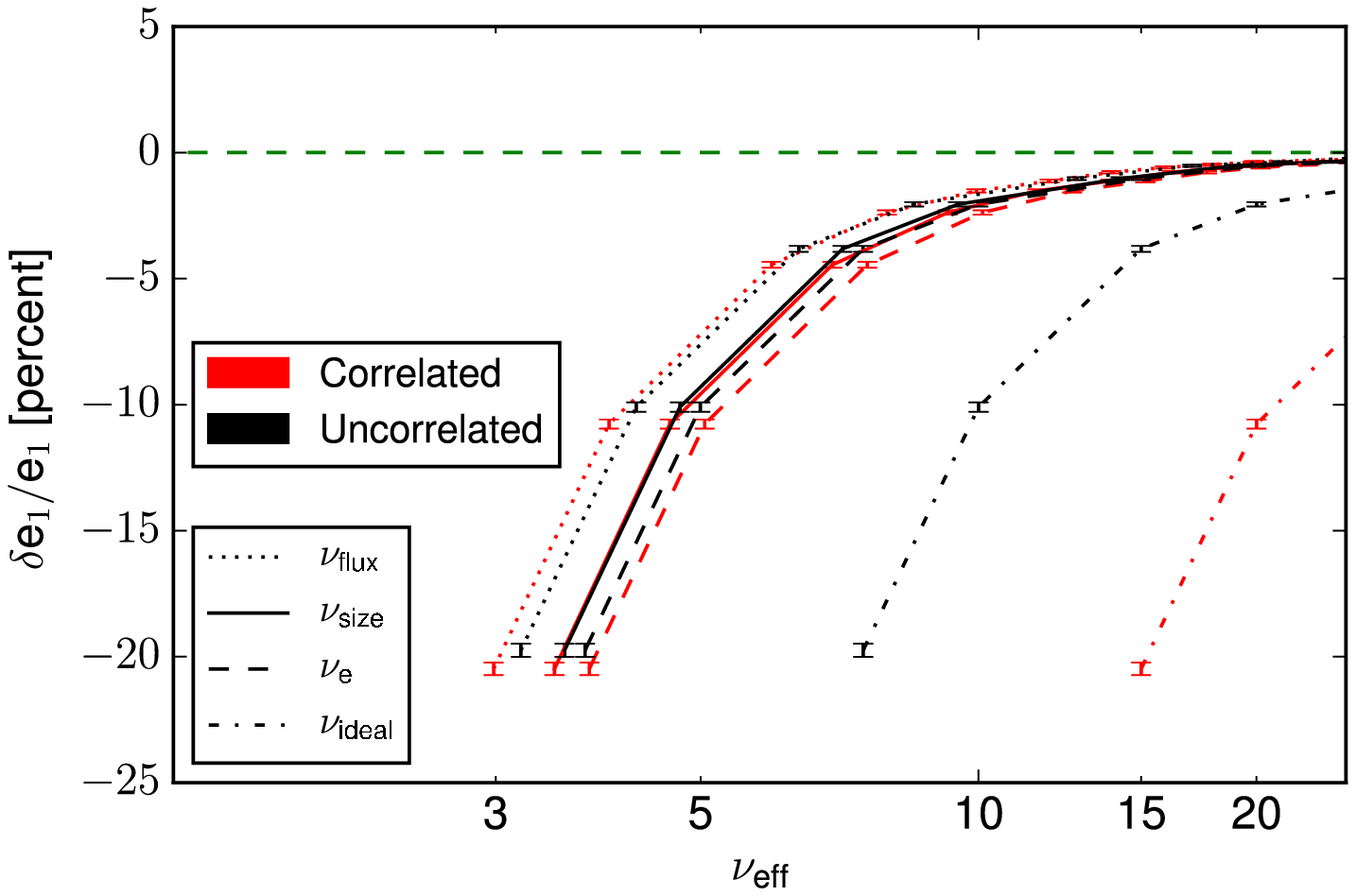}
	\caption{\label{F:space_6shape} 
Biases in shape measurements as a function of
      $\nu$, for different choices of \nuu{eff} (\nuu{e},
      \nuu{flux}, and \nuu{size}) and \nuu{ideal}, for both uncorrelated and correlated
      noise.  Unlike Fig.~\ref{F:6shape}, this plot shows results for space-like simulations using
      the parameters in Table~\ref{T:spaceparams}, for which the remapping by \nuu{eff} is also successful in 
	reducing the offset in the galaxy shape bias curves for correlated and uncorrelated 
	noise. }
	\end{center}
\end{figure}

Despite this difference in the error hierarchy, Fig.~\ref{F:space_6shape} clearly shows that our
conclusion about remapping shape biases using \nuu{eff} to account for the differences between
uncorrelated and correlated noise is also valid for space-like data.  In this case, the best
remapping is provided by \nuu{size}, while the worst comes from \nuu{e}.

\begin{table} 
		\begin{center}
			\begin{tabular}{|c | c |} 
				\hline $n$ & $1,2,3,4,5$ \\ \hline 
				$e_1$&$-0.45$, $0.45$\\ \hline 
				$e_2$ & $-0.45$, $0.45$\\ \hline 
			$r_{1/2}$ &$0.1\arcsec,0.2\arcsec,0.3\arcsec$\\ \hline 
				PSF &Airy ($\lambda/d$: .041253\arcsec) \\ \hline
				Pixel Scale & 0.03\arcsec / pixel \\ \hline
			\end{tabular} 
			\caption{
				\label{T:spaceparams}Parameters used for simulated space-based images.}
		\end{center} 
	\end{table}

\subsection{Generalizing to other methods}\label{subsec:generalizing}

An obvious question about the results earlier in this section is how specific are
they to re-Gaussianization, versus applying more generally.  For one galaxy model, the
same set of tests were carried out using \code{im3shape}, a maximum-likelihood forward model-fitting approach.  The configuration
settings used to run \code{im3shape} for this test are given in Appendix~\ref{app:im3shape}.
The galaxy model used for this test has the following parameters: $e=(0.45, 0.45)$, $n=2$,
$r_{1/2}=0.7$\arcsec, using our standard PSF and pixel scale for ground-based simulations from
Table~\ref{T:groundparams}.

\begin{figure} 
	\begin{center}
	\includegraphics[width=\columnwidth,angle=0]
	{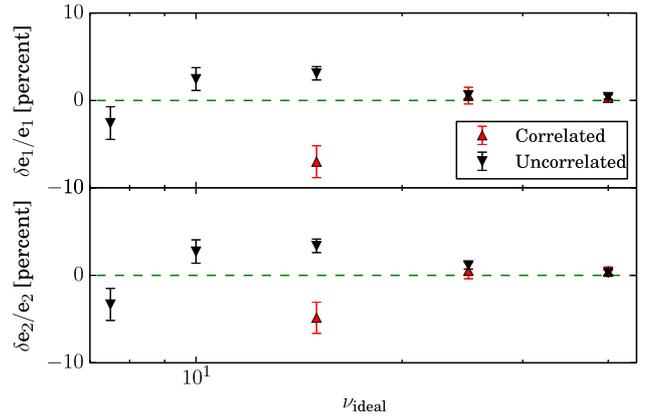}
	\caption{\label{F:im3shape} Galaxy shape bias for a single galaxy model as a function of
      \nuu{ideal} for \code{im3shape}, for uncorrelated and correlated noise models.}
	\end{center}
\end{figure}

First, the shape bias as a function of \nuu{ideal} is shown in Fig.~\ref{F:im3shape}.  
The results with uncorrelated noise can be compared against those from the upper left panel of figure 2 in
\cite{2012MNRAS.427.2711K}, which investigated noise bias in the context of \code{im3shape}.  While
the shape of our curve appears different from the one in \cite{2012MNRAS.427.2711K}, it is important
to keep in mind that our results go to lower signal-to-noise ratio.  For the range that the plots have in
common, the shape of the curve and magnitude of the bias at low detection significance is actually
fairly consistent (with some mild differences that could arise due to the galaxy models not being
identical).  The curve with correlated noise does not seem to go to positive values, unlike the
curve with uncorrelated noise; however, the curves have in common the behavior at the lowest
detection significance, where they both go to negative values.

\begin{figure*} 
	\begin{center}
	\includegraphics[width=6in]
	{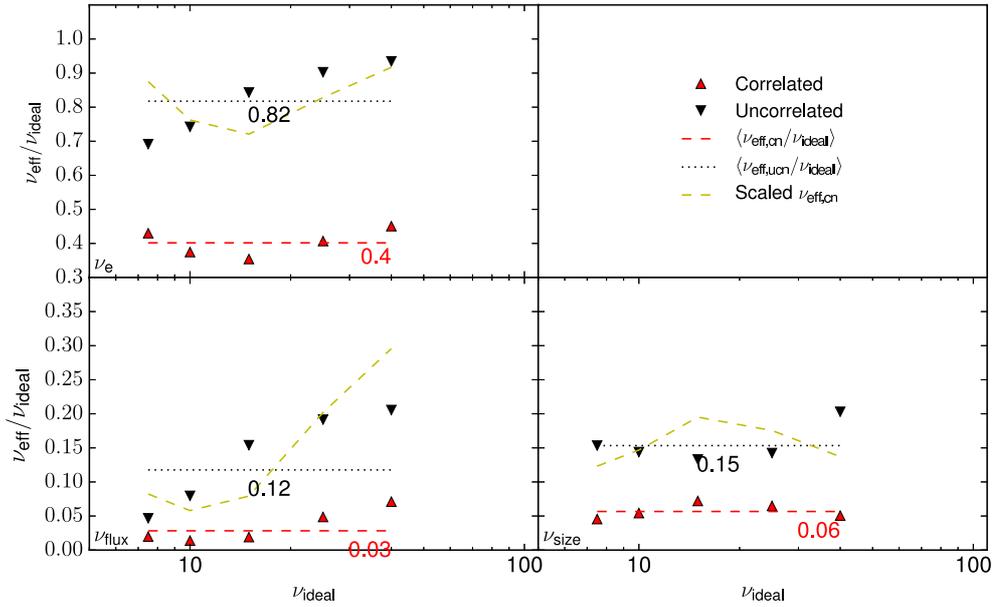}
	\caption{\label{F:im3_errors} 
Using \code{im3shape} measurements, \refresp{three} different measures of the 
	effective detection significance \nuu{eff} (\refresp{\nuu{e}, \nuu{size}, and \nuu{flux}}) are shown as $\nuu{eff}/\nuu{ideal}$ as a
    function of \nuu{ideal}.  The 
    average values of $\nuu{eff}/\nuu{ideal}$ across all \nuu{ideal}
    values is shown as $\langle\nuu{eff}/\nuu{ideal}\rangle$ (horizontal lines, with the
    average value given directly below each line).  
	A ``Scaled	\nuu{eff,ucn}'' line was constructed by multiplying the 
	red points (correlated noise) by the ratio of the
    $\langle\nuu{eff}/\nuu{ideal}\rangle$ lines for uncorrelated vs.\
    correlated noise. This allows for an easier comparison between the
  shapes of the curves for correlated and uncorrelated noise.
}
	\end{center}

\end{figure*}

\begin{figure} 
	\begin{center}
	\includegraphics[width=\columnwidth]
	{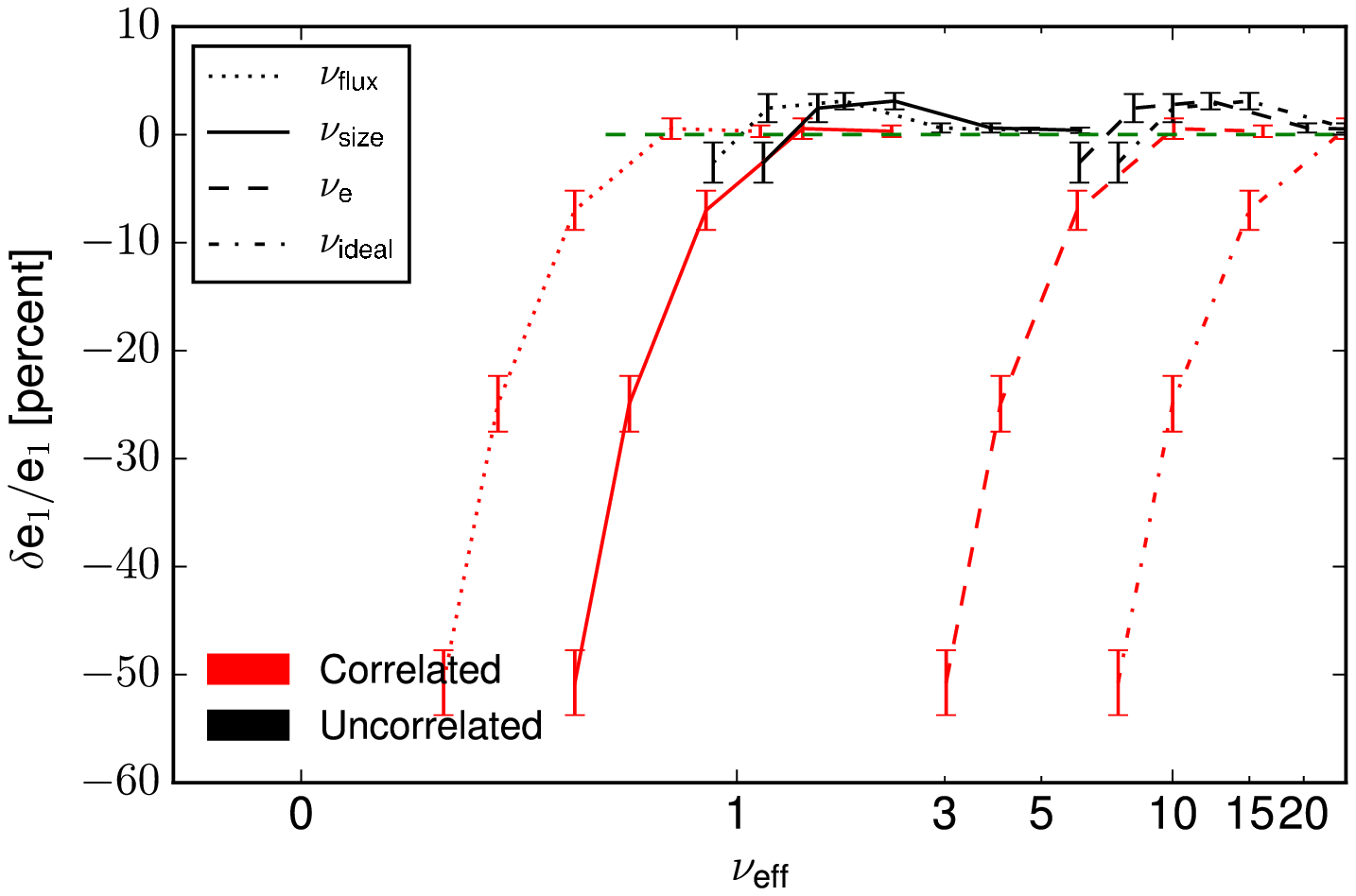}
	\caption{\label{F:im3_6shape}
Biases in shape measurements as a function of
      $\nu$, for different choices of \nuu{eff} (\nuu{e},
      \nuu{flux}, and \nuu{size}) and \nuu{ideal}, for both uncorrelated and correlated
      noise.  Unlike Fig.~\ref{F:6shape}, this plot shows results for \code{im3shape}, for which the
      remapping by \nuu{eff} is partially successful
	reducing the offset in the galaxy shape bias curves for correlated and uncorrelated 
	noise. 
  }
	\end{center}
\end{figure}

Next, we investigated the \nuu{eff} for \refresp{shape, size, and flux} from \code{im3shape}
(Fig.~\ref{F:im3_errors}).  It is important to bear in mind that \nuu{size} for
\code{im3shape} is intrinsically different from \nuu{size} from adaptive moments, since the former is
intrinsic size and the latter is PSF-convolved.  In general,
\nuu{size} from \code{im3shape} is therefore expected to be lower than that from adaptive moments.  Indeed,
Fig.~\ref{F:im3_errors} reveals quite low values of $\nuu{size}/\nuu{ideal}$.  \refresp{It is
  less clear why $\nuu{flux}/\nuu{ideal}$ is so low, implying that detections that are visually
  quite apparent in images have a flux-based detection significance around 1.}  Also, it is generally
the case that \refresp{$\nuu{eff}/\nuu{ideal}$ (for all three types of \nuu{eff})} are not consistent with being
flat, unlike the results in Fig.~\ref{F:errmap} from adaptive moments, though \nuu{size} is closer
to flat than \refresp{\nuu{e}}.  Nonetheless, our final step is to 
check the remapping of the galaxy shape biases from Fig.~\ref{F:im3shape} using \nuu{eff} in Fig.~\ref{F:im3_6shape}.  Clearly
the remapping brings the curves significantly closer together, though their behavior is not identical due to the
lack of a small positive bias for some range of detection significance when using correlated noise.
Thus, while the results for remapping from re-Gaussianization do not completely
carry over to \code{im3shape}, the remapping is still somewhat beneficial in reducing differences in
shape biases for uncorrelated and correlated noise.

\section{Conclusions}\label{sec:conclusions}

In this work, we have investigated the impact of correlated noise fields on galaxy shape estimation
using the re-Gaussianization galaxy shape estimation method.  Most galaxy shape
estimation methods ignore correlated noise, which may be present due to galaxies below the detection
threshold, or image resampling and 
other elements of the image processing.  Ignoring  correlated noise leads to overestimation of
the detection significance, and could in principle lead to incorrect conclusions about biases in
galaxy shape measurement.

Our results suggest that correlated noise tends to change the effective detection significance in
ways that depend only mildly on galaxy properties, and are easily characterized.  Moreover, while
the galaxy shape biases as a function of the variance of the noise field can differ dramatically for
uncorrelated and correlated noise, using an empirical detection significance estimator that takes
into account the impact of noise correlations can account for essentially all of this difference.
This statement is robust to significant changes in the parameters of the galaxies, PSFs, and
correlated noise fields in the images.  \refresp{While this result is perhaps unsurprising given the
  underlying cause of noise bias (as discussed in Sec.~\ref{sec:intro}), a direct validation using
  simulations is nonetheless valuable in confirming our understanding of noise bias.}

From a theoretical perspective, this result suggests that the correlated noise is
not doing anything untoward to the images, and the primary reason they change the results is due to the
degradation in detection significance for extended objects.  As a practical matter, this is useful;
 depending on the required level of accuracy, it may not be necessary to run suites of
simulations to calibrate noise bias for many different correlated noise fields, but rather that 
simulations with uncorrelated noise (or a single level of correlated noise) can be used to calibrate
noise bias even when the level of noise correlations is not completely uniform across the dataset.
This result rests on using a reliable estimator of detection significance that
includes the effect of noise correlations.  \refresp{The development of such a calibration scheme
 that properly takes all factors into account is beyond the scope of this work.}  In addition, the impact of anisotropy in the correlated
noise may cause shear biases that are significant enough to matter for large upcoming weak lensing
surveys when using galaxies near the detection limit.

Defining the detection significance can be non-trivial, and if it is done in a way that correlates
with shape, this can result in a selection bias with respect to the lensing shear
\citep[e.g.][]{2015arXiv150705603J}.  \refresp{While we have verified that shape measurement failure
  rates do not tend to depend on the magnitude of the galaxy ellipticity for either type of noise,
  it} is unclear whether the correlated noise could interact in some way with the definition of
detection significance in a way that modifies selection biases.  That issue is beyond the scope of
this paper.

Our results about the impact of correlated noise may explain some of the findings of
\cite{2015MNRAS.449..685H}, who note that the estimated shear calibration bias for galaxies limited
to $r$-band magnitude $<25$ depends on the limiting magnitude of the simulated galaxy sample, only
converging at a limiting magnitude fainter than 26.5.  The galaxies below the detection limit, while
not directly used for shear estimation, affect the results for the brighter galaxies that are used,
presumably due to the fact that they constitute a source of correlated noise.  If this is the
primary reason for the need to include galaxies below the detection limit in the simulations in
order to get the shear biases to converge, it is possible that the simulated galaxy sample could
actually be truncated slightly below (rather than well below) the detection limit if the noise field is generated with a level of noise
correlations consistent with the inclusion of fainter galaxies.  
Including the deblending issues that arise from objects right below the detection threshold
will require the simulated sample to go at least $\sim 0.5$ magnitudes fainter, but it is worth
investigating whether 
the rest of the effect could be incorporated using correlated noise.  This simplification would 
make the simulation process less expensive, since the steepness of the
galaxy number counts means that the sample of galaxies that must be simulated is
enlarged by a substantial factor when going $1.5$ magnitudes fainter. 

The generality of our result that shape biases can be remapped from uncorrelated to
correlated noise is unclear, though our preliminary results suggest that it is beneficial for one
other method of galaxy shape estimation \refresp{and mathematically there are reasons to expect it
  to be more widely valid}.   More generally, 
our results highlight the complexity of noise bias in the case of correlated noise, and
the need to account for its impact on effective detection significances when 
calibrating noise bias in weak lensing measurements.

\section{Acknowledgements} 

\refresp{We thank the anonymous referee for many useful suggestions that improved this paper.} 
We thank Barney Rowe and Joe Zuntz for their assistance in using \code{im3shape} for this work and
for useful conversations about the results, and Mike Jarvis for helpful comments on the paper itself.  RM
acknowledges the support of HST-AR-12857.01-A, provided by NASA through a grant from the Space
Telescope Science Institute, which is operated by the Association of Universities for Research in
Astronomy, Incorporated, under NASA contract NAS5-26555. 

\bibliographystyle{mnras}
\bibliography{papers}

\appendix 
\section{\code{im3shape} settings}\label{app:im3shape}

Table~\ref{T:im3shapeparams} shows the parameters used for all
\code{im3shape} plots used in this work.

\begin{table} 
		\begin{center}
			\begin{tabular}{cc} 
				\hline
				\textbf{Parameter} & \textbf{Value} \\
				\hline \hline
				\verb|model_name| & \verb|sersics| \\
				\verb|perform_pixel_integration| & \verb|N| \\
				\verb|verbosity| & \verb|-1| \\
				\verb|minimizer_verbosity| & \verb|-1| \\
				\verb|noise_sigma| & \verb|1.0| \\
				\verb|background_subtract| & \verb|N| \\
				\verb|upsampling| &          \verb|7| \\
				\verb|n_central_pixel_upsampling| &         \verb|0| \\
				\verb|n_central_pixels_to_upsample| &          \verb|0| \\
				\verb|padding| &          \verb|0| \\
				\verb|stamp_size| &         \verb|50| \\
				\verb|rescale_stamp| & \verb|Y| \\
				\verb|minimizer_max_iterations| &        \verb|150| \\
				\verb|levmar_eps1| &      \verb|1e-40| \\
				\verb|levmar_eps2| &      \verb|1e-40| \\
				\verb|levmar_eps3| &      \verb|1e-40| \\
				\verb|levmar_eps4| &      \verb|-1| \\
				\verb|levmar_eps5| &      \verb|1e-05| \\
				\verb|levmar_tau| &      \verb|1e-5| \\
				\verb|levmar_LM_INIT_MU|&      \verb|1e-20| \\
				\verb|minimizer_loops| &          \verb|1| \\
				\verb|sersics_bulge_A_min| &        \verb|-50| \\
				\verb|sersics_radius_start| &         \verb|10| \\
				\verb|psf_input| &\verb|psf_image_single| \\
				 \hline
			\end{tabular} 			
			\caption{
				\label{T:im3shapeparams}Parameters used for \code{im3shape}.}
		\end{center} 
	\end{table}

\end{document}